\documentclass{PoS}

\title{Three Dimensional Simulations of the Core Helium Flash - with Rotation}

\ShortTitle{3D Core He Flash}

\author{\speaker{John Lattanzio}\\
        Centre for Stellar and Planetary Astrophysics, Monash University\\
        E-mail: \email{john.lattanzio@sci.monash.edu.au}}

\author{David Dearborn\\
        Lawrence Livermore National Laboratory\\
        E-mail: \email{dearborn2@llnl.gov}}

\author{Peter Eggleton\\
        Lawrence Livermore National Laboratory\\
        E-mail: \email{ppe@igpp.ucllnl.org}}

\author{Don Dossa\\
        Lawrence Livermore National Laboratory\\
        E-mail: \email{dossa1@llnl.gov}}

\abstract{We continue our study of the core helium flash using the three 
dimensional hydrodynamics code {\it Djehuty\/}. Continuing from earlier
calculations, we now take relaxed 3D configurations and add various amounts of
rotation. We find that rotation periods consistent with those observed in
white dwarfs produce negligible changes in the structure and evolution
of the core flash, at least for the very small timescales we have yet
been able to investigate. There is no sign of any extra mixing due to
the rotation. There is some inconclusive evidence for a slight
change in the luminosity, at the 1\% level.}

\FullConference{International Symposium on Nuclear Astrophysics --- Nuclei in the Cosmos --- IX\\
		 June 25-30 2006\\
		 CERN, Geneva, Switzerland}

\begin{document}

\section{Introduction}
The core Helium flash remains one of the most challenging phases for
computational stellar astrophysics, at least for low mass stars. The
combination of high degeneracy and highly temperature dependent reactions 
provides for near-explosive behaviour, and yet observations show that
the stars survive this phase to continue life on the horizontal branch, 
burning He in a quiescent manner. Many of the assumptions that are 
common in spherically symmetric hydrostatic codes may well be violated 
during the core flash: hydrostatic equilibrium itself being one such
assumption, spherical symmetry another, and instantaneous mixing being 
yet another. Even if a diffusion equation is used for mixing, the
timescale for change is so short that it casts doubt on the applicability of
that regime. Indeed, all convection models, and associated energy transport,
are highly suspect under such conditions. There is no substitute for direct
hydrodynamical calculations of this phase.

\section{Previous Multi-Dimensional Calculations}
There have been precious few calculations of the core flash in more 
than one dimension, reflecting the computational 
demands of such a project. The pioneering work of Bob Deupree and Peter Cole 
(\cite{D96} and references therein) suffered from the computational
limitations of the day, but nevertheless made some important discoveries.
Perhaps the most important was that there was significant overshoot
of the convective regions inwards, mixing closer to the core than found in
the 1D models. This has important implications for models of novae, for
example (see the paper by Jordi Jose, at this meeting). It may also
mean that the mini-pulses seen in the 1D models, which are required to
burn the He all the way into the centre of the star, do not in fact develop
in real stars. 

Our own previous calculations of the core flash with the {\it Djehuty\/} 
code also confirmed this (\cite{DLE06}). Over the time of our calculations we
found that the 3D models reproduced quite accurately the structure 
found in the 1D model. Sonic perturbations resulting from the mapping 
from 1D to 3D result in large oscillations in the stars and various 
amounts of damping are required to try to remove these temporary 
transients. We have not (yet) been able to run our calculations far 
enough to verify that the maximum extent of the convective motions 
in the 3D code match those found in the 1D calculation, but we did
verify the overshoot inwards previously found by Deupree (see Figure~1). We  
found that there was no tendency for deviations from spherical symmetry, 
even in the case where artificial and large temperature perturbations 
were applied. There is also a tendency for the 3D model to show 
luminosities that are about a factor of two lower, but this could 
be the result of many effects, such as the transients associated with
the mapping to 3D or the start of convective motions, coupled with
the very high temperature dependence of the triple alpha reaction.
Recall that initially the luminosity jumps briefly, due to
exactly these effects, and this initiates an expansion which we would
expect to result in a slightly lower energy output until the configuration
relaxes back to its stable state.
Further calculations are required to clarify this.

\begin{figure}
\centerline{\includegraphics[width=.7\textwidth]{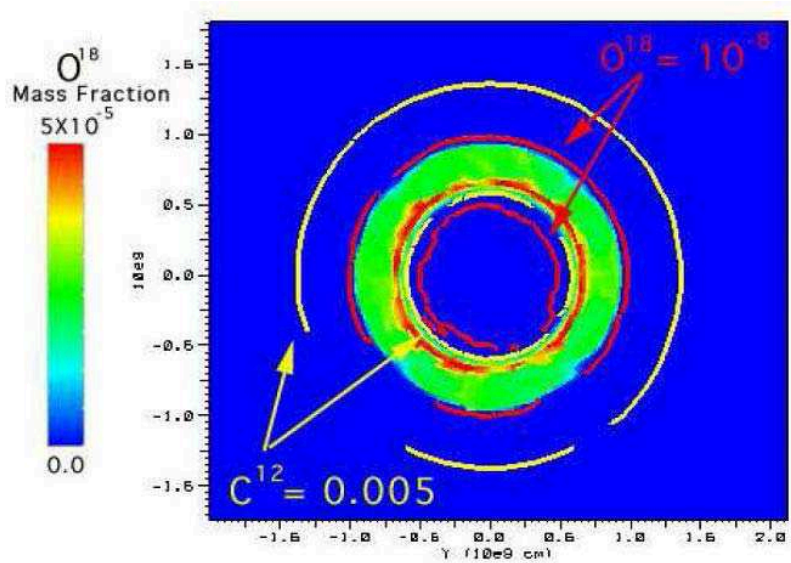}}
\caption{Mixing contours on a slice through the centre of a non-rotating
model, approximately one hour since the beginning of the calculation. 
The extent of convection in the 1D case is shown by the yellow 
contours, and the maximum extent of mixing found in our model, so far, is 
shown in red. These $^{18}$O contours show the extent of mixing because
this species is not included in the 1D code used to start the calculations,
but the formation of $^{18}$O from alpha capture on $^{14}$N in the 3D code
shows the extent of the mixing.}
\label{fig1}
\end{figure}

\section{The {\it Djehuty\/} Code}
We make use of the {\it Djehuty\/} hydrodynamical code developed at 
Lawrence Livermore National Laboratory. This is an explicit 3D code, using the
ALE (Arbitrary Lagrange-Eulerian) mesh scheme, which is second order in both
space and time. It includes all of the essential physics usually seen
in codes for calculating stellar interiors: detailed equation of state,
opacities covering the appropriate temperature range
(including low temperatures and conductive opacities),
a suite of some 21 nuclear  species, radiative diffusion, and 
self-gravity (assumed spherically symmetric at present). Note that
this assumption would be expected to slow the growth of any
non-spherical perturbations. We do not see any form, but we are aware that
this is a limitation of the present implementation of the code.

Our initial model comes from a 1D calculation that uses the same input physics,
thus minimizing the transients associated with a change of code. Nevertheless,
some perturbations introduced necessarily in the transfer to a 3D grid. 
These generate sonic waves that can bounce around the simulation for quite
some time. Further, although the 1D code tells us where convection is
expected (at least, according tot he Schwarzschild criterion) this does not help 
us in initializing the convective regions in the fully hydrodynamical code.
In the latter case, motions are driven by convective buoyancy and we must wait
for the code to determine the resulting flow patterns. This also leads to
transients which must be damped, using the technique discussed in \cite{DLE06}.

\section{Adding Rotation}
To begin our investigation of the effects of rotation, we took one of
the non-rotating models near the peak of the He flash (E9 in the terminology
of \cite{DLE06}); this model had a luminosity of approximately 
$3\times 10^9$L$_\odot$
in the 1D case, and had been taken forward some 3660 second in the
full 3D code. At this stage the luminosity had settled down to about 
$1.7\times 10^9$L$_\odot$. At this stage we introduce solid body 
rotation. Of course, even if we believe the rotation to be solid-body at the
time of the core-flash, we would expect some modifications to that
soon after. We ignore these for now.
We use three different periods: Spin1 with a period of 4.8 hours,
Spin2 with a period of 1.6 and Spin3 with a period of 0.87 hours. These
cover the periods  typically seen in white dwarfs \cite{Spruit}. As seen in Figure~2, 
the centrifugal acceleration thus introduced is still a tiny fraction
of the gravitational acceleration, at least in the core of the star.
The rotational velocities at the edge of the core are of order 2, 6 and 12 km/sec respectively, and this should be compared with the typical convective motion speeds of about 40 km/sec. All of these considerations lead us to predict that rotation will not have a dramatic effect on the structure of the core flash. (Sadly, we cannot take the calculations far enough to discuss the {\it evolution\/} of the core flash.)

\begin{figure}
\centerline{\includegraphics[width=.8\textwidth]{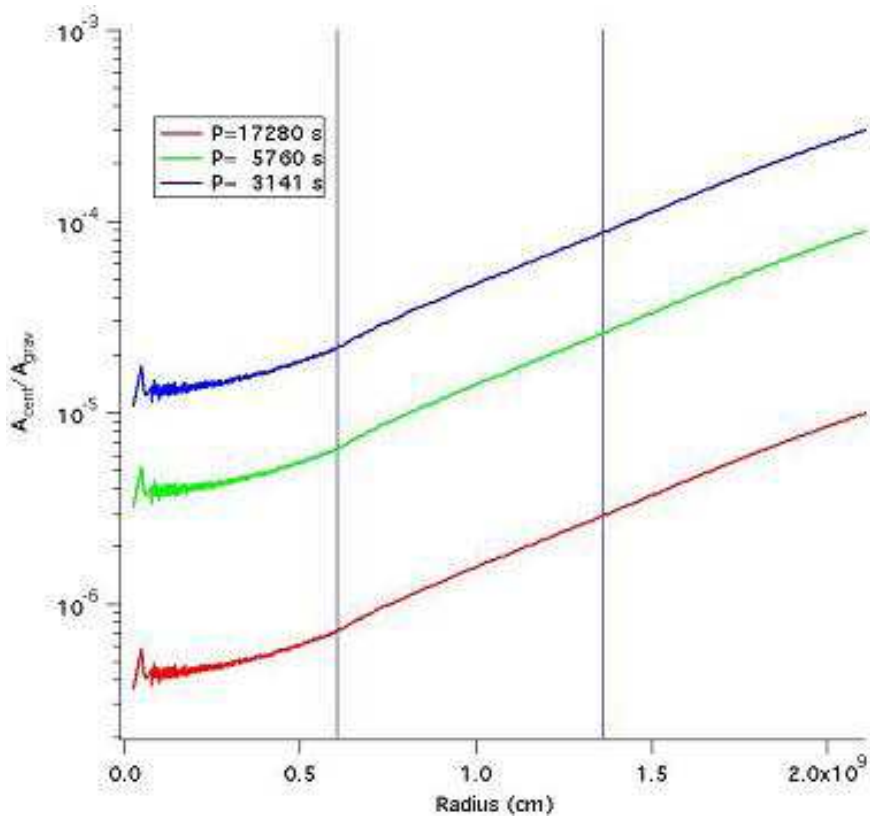}}
\caption{The ratio of the centrifugal acceleration to the gravitational
acceleration in the three rotating cases. The two vertical lines show the
extent of convection in the 1D model.}
\label{fig2}
\end{figure}

Initially, the Spin1 case is indistinguishable from the non-rotating case.
But gradually it settles down to a slightly higher luminosity. 
We speculate that this is due to the limited ability of convection to
transport energy vertically in the rotating case. As a plume rises, it begins 
to curve to the side from Coriolis forces.  As a result it
must penetrate more surrounding material to reach the same height.  
We believe that this reduces the rate at which energy is spread from the 
very thin burning region at the bottom, resulting in a somewhat higher luminosity.
After
some 2000 seconds of star time the luminosity is 0.50\% higher than in the non-rotating case, although the fluctuations are typically 0.36\%.
Case Spin2 increased its luminosity also, but by only 0.29\% whereas 
Spin3 decreased by a tiny 0.03\%. We are not yet convinced of any systematic
trend with rotation rate.

Figure~3 shows how the speed varies with radius for Spin3. The initial
imposed rotation profile is shown, together with the average over four (only!)
rays from the centre. One can see that the motion is largely following the
initially imposed rotation except in the convective regions where the 
velocity magnitude is independent of radius, and also shows that
convection has extended inwards, eating into the core, as found by Deupree
and co-workers as well as in our earlier calculations.

\begin{figure}
\centerline{\includegraphics[width=.8\textwidth]{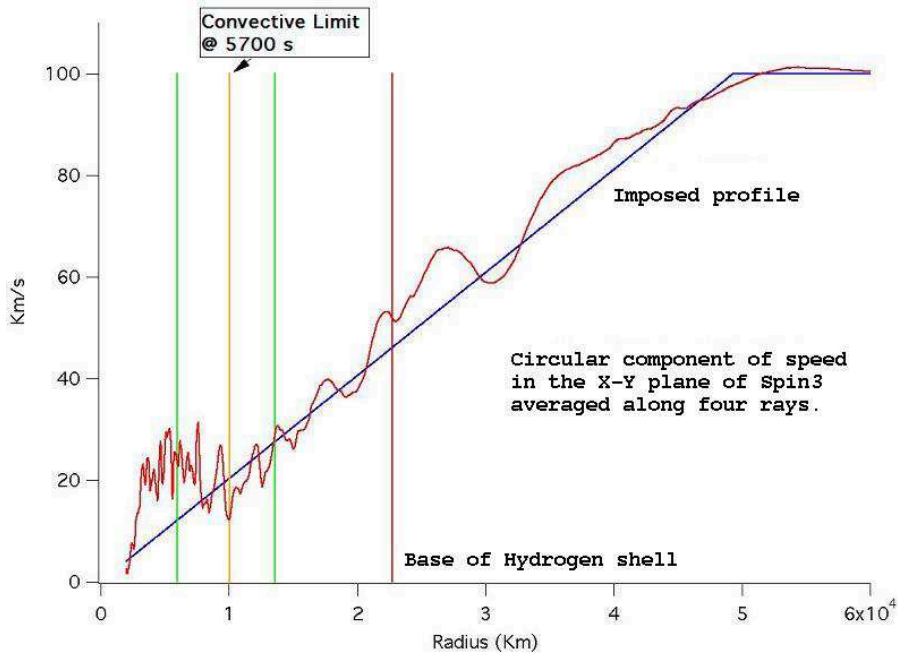}}
\caption{Circular component of the speed in the x-y plane of Spin3. The initial imposed profile is shown in the blue line, and the red is the average along four radial rays. The green vertical lines show the limits of convection in the 1D case, and the yellow line shows the maximum extent of the convection so far in the 3D calculation. The vertical red line shows where the H-shell is located.}
\label{fig3}
\end{figure}

Our final figure shows a contour of $^{18}$O from the Spin3 run at the
level of $4\times 10^{-4}$. The 1D code used for the initial configuration 
does not include $^{18}$O so there is none in the simulation initially.
However, the 3D code allows for alpha captures on $^{14}$N so the resultant $^{18}$O is a good tracer of the edge of the helium burning region. Figure~4 shows one $^{18}$O convective "plume" that has risen and then twisted as a result of the rotation in the core. The shear stops the plume from rising very far, at this stage. We see many such plumes, typically rising 3000km in about a minute.

The growth of the extent of the convective regions is very similar to 
the case when rotation is not present. For example, the case E9 saw convection extend by 451km in 2000 seconds, and Spin1 grew by 458km over the same time.
The outer boundary of the convection is a little less smooth in the rotating case, but there is, as yet, no evidence for enhanced mixing,. Further calculations are needed to expand on this point, and will be the subject of a separate publication.

\begin{figure}
\centerline{\includegraphics[width=.8\textwidth]{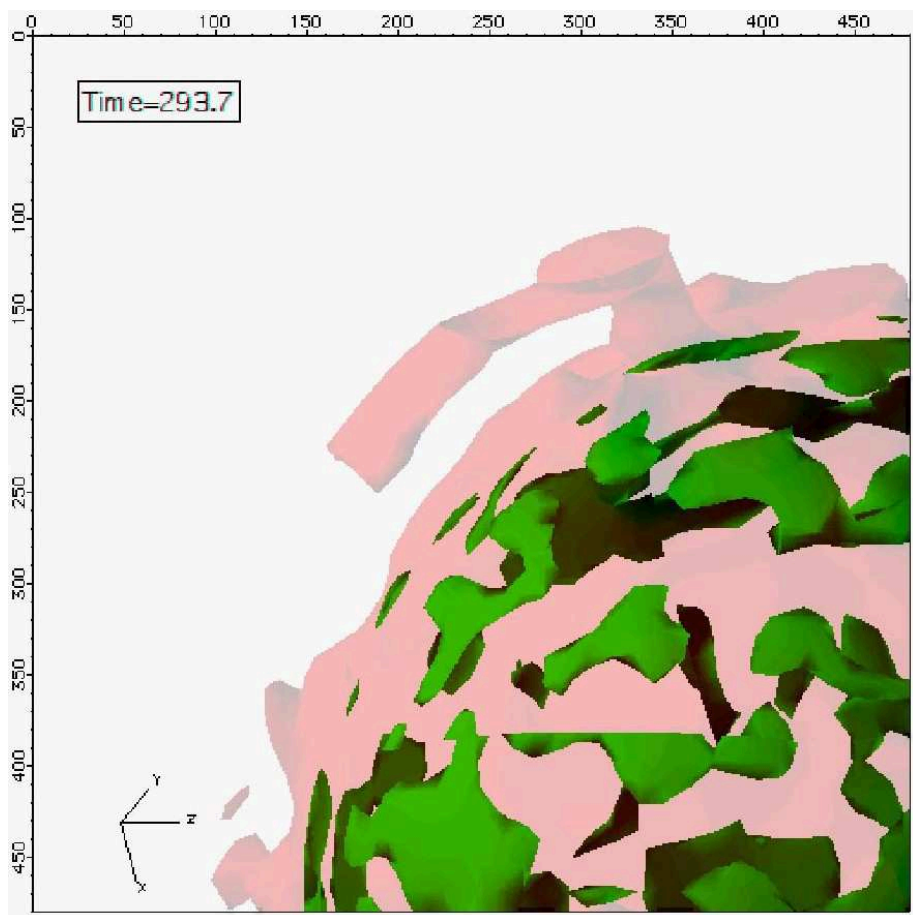}}
\caption{Pink levels show the contour where $^{18}$O is $4\times 10^{-4}$ and the green layers are where the temperature is 188 million K.}
\label{fig4}
\end{figure}

\section{Conclusions}
We have presented some preliminary results from an investigation of 
rotating models of 3D core helium flash simulations. We imposed solid-body rotation profiles at rates consistent with the fastest observed white dwarf rotation periods. The effects of such rotation rates are minimal, at least on the
helium burning and convecting region, and over the short timescales covered by our calculations (approximately 2000 seconds).

We see sort-lived hot-spots and rising convective plumes, but no evidence
(yet) for enhanced mixing. There is a slight effect on the luminosity, 
of order 1\%, but no consistent trend is observed.

Further calculations are in progress.

\section*{Acknowledgments}
This study has been carried out under the auspices of the
U.S. Department of Energy, National Nuclear Security Administration, 
by the University of California, Lawrence Livermore National 
Laboratory, under contract
No.~W-7405-Eng-48, and was partially supported by the 
Australian Research Council.

\end{document}